\documentclass[aps,prb,twocolumn,superscriptaddress,showpacs,floatfix]{revtex4}

\newcommand{\vrr}{\mathbf{r}}
\newcommand{\vrro}{\mathbf{r}_{1}}
\newcommand{\vrrt}{\mathbf{r}_{2}}
\newcommand{\intr}{\int \! d\vrr\,}
\newcommand{\introt}{\int \! d\vrro \, d\vrrt\,}
\newcommand{\psir}{\psi (\vrr)}
\newcommand{\psiro}{\psi (\vrro)}
\newcommand{\psirt}{\psi (\vrrt)}
\newcommand{\prob}[1]{\left|#1\right|^2}
\newcommand{\ppsir}{\prob{\psir}}
\newcommand{\ppsiro}{\prob{\psiro}}
\newcommand{\ppsirt}{\prob{\psirt}}
\newcommand{\dpsi}{\prob{\nabla\psir}}
\newcommand{\dist}[1]{D(#1)}
\newcommand{\disto}[1]{D_{0}(#1)}
\newcommand{\eps}{\epsilon_{s}}
\newcommand{\epinf}{\epsilon_{\infty}}
\newcommand{\epstar}{\epsilon^{*}}
\newcommand{\bind}[1]{E_{b}^{\mathrm{#1}}}
\newcommand{\bexc}{\bind{exc}}
\newcommand{\bpol}{\bind{pol}}
\newcommand{\bindo}[1]{E_{0}^{\mathrm{#1}}}
\newcommand{\bexco}{\bindo{exc}}
\newcommand{\bpolo}{\bindo{pol}}
\newcommand{\Jfmin}[1]{\mathcal{E}^{\mathrm{#1}}\{\psi \}}
\newcommand{\Jfexc}{\Jfmin{exc}}
\newcommand{\Jfpol}{\Jfmin{pol}}
\newcommand{\Jmin}[1]{\mathcal{E}^{\mathrm{#1}}}
\newcommand{\Jexc}{\Jmin{exc}}
\newcommand{\Jpol}{\Jmin{pol}}
\newcommand{\Kin}[1]{K^{\mathrm{#1}}}
\newcommand{\Pot}[1]{U^{\mathrm{#1}}}
\newcommand{\Kexc}{\Kin{exc}}
\newcommand{\Uexc}{\Pot{exc}}
\newcommand{\Kpol}{\Kin{pol}}
\newcommand{\Upol}{\Pot{pol}}
\newcommand{\Jfmino}[1]{\mathcal{E}^{\mathrm{#1}}_{0}\{\psi \}}
\newcommand{\Jfexco}{\Jfmino{exc}}
\newcommand{\Jfpolo}{\Jfmino{pol}}
\newcommand{\Jmino}[1]{\mathcal{E}^{\mathrm{#1}}_{0}}
\newcommand{\Jexco}{\Jmino{exc}}
\newcommand{\Jpolo}{\Jmino{pol}}
\newcommand{\aB}{a_{0}}
\newcommand{\Ro}{R_{0}}

\newcommand{\Ry}{\mathrm{Ry}}
\newcommand{\Rys}[1]{\Ry^{\mathrm{#1}}}
\newcommand{\al}{\alpha}
\newcommand{\be}{\beta}
\newcommand{\ga}{\gamma}
\newcommand{\yo}{y_{1}}
\newcommand{\Veff}{V_{\mathrm{eff}}(x)}

\usepackage{graphicx}
\usepackage{amsmath}
\usepackage{latexsym}
\begin{document}

\title{Nanotubes in polar media:
polarons and excitons on a cylinder}

\author{Yu.~N.~Gartstein}
\affiliation{Department of Physics, The University of Texas at
Dallas, P. O. Box 830688, FO23, Richardson, Texas 75083, USA}
\author{T.~D.~Bustamante}
\affiliation{Department of Physics, The University of Texas at
Dallas, P. O. Box 830688, FO23, Richardson, Texas 75083, USA}
\author{S.~Ortega Castillo}
\affiliation{FAMAT, University of Guanajuato, Guanajuato, Mexico}

\date{\today}

\begin{abstract}
Electrons and holes on semiconducting nanotubes immersed in
sluggish polar media can undergo self-localization into polaronic
states. We evaluate the binding energy $\bpol$ of adiabatic
Fr\"{o}hlich-Pekar polarons confined to a cylindrical surface and
compare it to the corresponding exciton binding energy $\bexc$.
The ratio $\bpol/\bexc$ is found to be a non-monotonic function of
the cylinder radius $R$ that can reach values of about $0.35$,
substantially larger than values of about $0.2$ for 2$d$ or 3$d$
systems. We argue that these findings represent a more general
crossover effect that could manifest itself in other semiconductor
nanostructures in 3$d$ polar environments. As a result of the
strong polaronic effect, the activation energy of exciton
dissociation into polaron pairs is significantly reduced which may
lead to enhanced charge separation.
\end{abstract}


\maketitle

\section{Introduction}

Low-dimensional semiconductor structures such as quantum wells,
quantum wires, nanotubes and conjugated polymers are important for
practical applications and interesting scientifically. It is known
that the confinement of the motion of charge carriers in some
directions leads to increased effects of the Coulomb interaction
on system excitations. In this paper we are concerned with two
types of such effects: \textit{excitonic} and \textit{polaronic}.
The excitonic effect refers to the formation of Coulombically
bound electron-hole pairs, Wannier-Mott excitons, which
progressively affect optical properties of semiconductors: the
exciton binding energy $\bexc$ increases from its 3$d$ value to
2$d$ and, further on, to quasi-1$d$ magnitudes ($\bexc$ diverges
in pure 1$d$).\cite{haugbook} The polaronic effect occurs in polar
media, where the Coulomb field of an individual charge carrier
causes the polarization (deformation) of its surroundings
resulting in the carrier self-localization into polaronic states.
Such polarons have been especially extensively studied in the
context of 3$d$ ionic crystals and polar
semiconductors.\cite{froreview,polarons1,appel} The polaronic
effect also grows with the confinement: the polaron binding energy
$\bpol$ in 2$d$ is larger than in 3$d$ and would diverge in pure
1$d$.\cite{pol2d,pol1d} Various aspects of the excitonic and
polaronic effects have been explored for many specific
low-dimensional systems.

Of particular interest to us is a relationship between $\bexc$ and
$\bpol$, each of the binding energies understood as measured from
the band edges in the absence of the polaronic effect. The ratio
$\bpol/\bexc$ of the binding energies has a clear significance for
the relative energetics of closely-bound and well-separated
electron-hole pairs that is expected to affect practically
important processes of charge separation and recombination. The
bare value $\bexc$ signifies the ionization energy, whether
thermal or photo, required for ``unbinding'' of the exciton into
well-separated band-edge electron and hole. In the presence of the
polaronic effect, however, the thermal ionization (dissociation)
would occur into distant electron-polaron and hole-polaron so that
the exciton thermal ionization  energy is reduced from $\bexc$ to
$\bexc-2\bpol$. We are interested in how much of this relative
reduction might be possible to achieve due to the formation of
\textit{strong-coupling} (adiabatic) polarons. Our discussion here
is restricted to systems with equal electron and hole effective
masses so that the electron-polaron and hole-polaron have the same
binding energies while the neutral exciton does not experience the
adiabatic polaronic energy correction.\cite{pekar2,rashba}

It is instructive to look at the ratio $\bpol/\bexc$ based on the
results known for isotropic systems of ``well-defined''
dimensionality, that is, for purely 3$d$, 2$d$ and 1$d$ systems.
One would then find that, while each of the binding energies
increases with more confinement, their growth occurs nearly ``in
proportion'' so that the ratio does not change significantly.
Indeed, classic Pekar's result\cite{pekar1} for 3$d$ adiabatic
polarons would translate into the maximum ratio of about 0.22. The
exciton binding in 2$d$ increases by a factor of 4 from its 3$d$
value\cite{haugbook} but almost so does the polaron binding energy
in 2$d$\cite{pol2d} resulting in the ratio $\simeq 0.20$.
Moreover, if the divergent purely 1$d$ binding energies are taken
(parametrically) for their ratio, then the result of
Ref.~\onlinecite{pol1d} would translate into maximum $\bpol/\bexc$
of approximately 0.17.

From the standpoint of this data, the results of our recent
quasi-1$d$ calculations\cite{YNGpol} for polarons and excitons on
nanotubes immersed in a 3$d$ polar medium yielding $\bpol/\bexc$
in excess of 0.3 appear quite surprising. In the particular case
of the tubular geometry, charge carriers are confined to the
motion on a cylindrical surface. Our study\cite{YNGpol} was
restricted to relatively small cylinder radii $R$. In this paper
we will use a direct variational approach to calculate adiabatic
polarons for arbitrary $R$ thereby enabling an assessment of the
evolution of the ratio $\bpol/\bexc$ between the purely 2$d$ ($R
\rightarrow \infty$) and quasi-1$d$ regimes. The corresponding
calculations for excitons on a cylindrical surface have been
performed recently,\cite{kostov,pedersen} which are very much in
line with our exciton data. We will explicitly demonstrate (see
Fig.~\ref{Energies}(b)) that the ratio of the polaron and exciton
binding energies exhibits a non-monotonic behavior as a function
of the cylinder radius and can achieve as large values as about
0.35 at intermediate $R$ where the cylinder circumference is
roughly comparable to an appropriate Bohr radius. We believe that
our demonstration of $\bpol/\bexc$ ratios above the values in
purely 3$d$ and 2$d$ systems can be rationalized by invoking
simple physical arguments. These arguments also suggest that the
found \textit{relative} increase of the polaronic effect may be
reflective of a general behavior that might be characterized as a
crossover effect.

Consider a gradual increase of the confinement, e.~g., by starting
to decrease the radius of a very large cylinder in going from the
purely 2$d$ system towards quasi-1$d$ or by starting to decrease
the thickness of a very thick quantum well in going from the
purely 3$d$ system towards 2$d$. Simplistically stated, the
spatial size (extent) of a yet unconfined polaron wave function is
larger than that of an unconfined exciton. As the confinement
increases, therefore, the polaron can start experiencing a
substantial growth of its binding energy due to the confinement
before a ``proportionally'' substantial growth of the exciton
binding energy sets in. With still further increase of the
confinement, both polaron and exciton binding energies will be
reflecting fuller confinement effects resulting in the
corresponding trend of the decreasing ratio $\bpol/\bexc$. Of
course, quantitative aspects of the evolution can vary for
different systems and need to be evaluated accordingly. We also
note that the above consideration tacitly assumed the existence of
a uniform 3$d$ polarizable medium with the confinement affecting
only the motion of charge carriers. Strong violations of this
assumption can significantly affect the outcome for the ratio of
binding energies.

While serving as a suggestive illustration of a possibly general
behavior, it is polarons and the relationship of polarons and
excitons on a cylindrical surface that are the subjects of our
direct interest in this paper. Organic and inorganic tubular
(nano)structures attract a great deal of attention and are
considered candidate systems for important applications like
(photo)electrochemical energy conversion and (photo)catalytic
production and storage of hydrogen as well as in optoelectronics.
On one hand, their extended size along the tube axis can
facilitate very good electron transport in that direction. On the
other hand, tubes can expose large areas of both exterior and
interior surfaces to facilitate surface-dependent reactions. Many
of these applications involve contact with polar liquid
environments such as common aqueous and non-aqueous solvents and
electrolytic solutions which can provide conditions appropriate
for the strong polaronic effect\cite{YNGpol} thereby changing the
nature of charge carriers. A widely known example of the tubular
geometry is single-walled carbon nanotubes (SWNTs) and, in fact,
``chiral selective reactivity and redox chemistry of carbon
nanotubes are two emerging fields of nanoscience''.\cite{chirsel}
We note that the importance of the excitonic effects in the optics
of semiconducting SWNTs is well established now with the radius
$R$-dependent binding energies $\bexc$ experimentally measured in
some SWNTs to be in the range of $0.4 - 0.6$
eV.\cite{optrescnt,bachilo_bind,krauss1} There is also a growing
evidence of the environmental effects on the electronic properties
of SWNTs.\cite{fantini,hertel,mplu} The assumption of equal
electron and hole effective masses is a good approximation for
SWNTs.

It can be anticipated that the polaronic effect would have an
influence both on charge-transfer reactions and charge carrier
dynamics on the tubes. In the context of our discussion of the
relative energetics, a substantial reduction of the activation
energy due to the polaronic effect should be expected for
electric-field-assisted exciton dissociation and charge separation
on semiconducting nanotubes.

\section{Exciton and polaron energy functionals}

Following Refs.~\onlinecite{kostov} and \onlinecite{pedersen}, the
basic model we consider in this paper assumes that electron and
hole are particles of the same effective mass $m_{e}=m_{h}=m$
whose motion is restricted to the surface of the cylinder of
radius $R$. The cylindrical surface itself is immersed in the
uniform 3$d$ dielectric continuum characterized, as is common in
studies of the polaron\cite{froreview,polarons1,appel} and
solvation\cite{fawcett} effects, by two magnitudes of the
dielectric permittivity: the high-frequency (optical) value of
$\epinf$ and the low-frequency (static) value of $\eps$. In the
case of liquid polar media, the slow component of the polarization
is ordinarily associated with the orientational polarization of
the solvent dipoles and it is typical\cite{fawcett,YNGpol} that
$\eps \gg \epinf$. The fast component of the polarization follows
charge carriers instantaneously. The slow component of the
polarization, on the other hand, is considered static in the
adiabatic picture to determine the electronic states; the slow
component then responds to the averaged electronic charge
distribution.

With $m_{e}=m_{h}$, there is no net charge density associated with
the ground state of the neutral exciton that would cause a static
polarization of the slow component (``non-polarizing exciton'').
The Coulomb interaction between the electron and the hole in the
exciton is screened only by the high-frequency dielectric response
$\epinf$.\cite{pekar2,rashba,pekar3} The total energy of the
exciton as a functional of the normalized wave function $\psir$ of
the electron-hole relative motion (reduced mass $m/2$) is then
\begin{eqnarray}
\Jfexc & = & \Kexc - \Uexc \nonumber \\
& = & \frac{\hbar^2}{m}\intr \dpsi - \frac{e^2}{\epinf}\intr
\frac{\ppsir}{\dist{\vrr}}, \ \ \ \ \ \label{Jexc1}
\end{eqnarray}
whose global minimum determines the exciton ground state wave
function and its binding energy. For our geometry, position vector
$\vrr=(x,y)$ is confined to the cylindrical surface, where we
choose $x$ to be along the cylinder axis and $-\pi R < y < \pi R$
 along the circumferential direction;
$\dpsi=(\partial\psi/\partial x)^2 + (\partial\psi/\partial y)^2$.
Coulomb interaction is determined by the physical distance in the
3$d$ space; in the flat geometry it would be $\dist{\vrr}=|\vrr|$,
for the points on the cylindrical surface
\begin{equation}\label{dist}
\dist{\vrr}=\left(x^2 + 4R^2 \sin^2\frac{y}{2R}\right)^{1/2}.
\end{equation}
Optimization of the functional (\ref{Jexc1}) for the ground state
of the exciton on a cylinder has been performed in
Refs.~\onlinecite{kostov} and \onlinecite{pedersen} with the
results in a very good agreement with our data to be used in the
comparison with the polaron.

Polarons we discuss here are of the large-radius
Fr\"{o}hlich-Pekar type where the Coulomb field of an individual
charge carrier supports a self-consistent dielectric polarization
pattern surrounding the carrier.  The formation of large-radius
polarons (self-localization, self-trapping) occurs due to the
interaction of charge carriers with the ``slow'' component of
polarization. As the fast component of polarization does not
contribute to the polaronic effect, it is the effective dielectric
constant $\epstar$:
\begin{equation}\label{epstar}
1/\epstar=1/\epinf  - 1/\eps,
\end{equation}
that affects the coupling
strength.\cite{froreview,polarons1,appel} In the adiabatic
approximation, the normalized wave functions $\psir$ of a
self-localized charge carrier correspond to the minima of the
following polaron energy functional:
\begin{eqnarray}
\Jfpol & = & \Kpol - \Upol \nonumber \\
& = & \frac{\hbar^2}{2m}\intr  \dpsi \nonumber \\
& - & \frac{e^2}{2\epstar}\introt \frac{\ppsiro \,
\ppsirt}{\dist{\vrro-\vrrt}}, \ \ \label{Jpol1}
\end{eqnarray}
whose global minimum we will be seeking for the ground state of
the polaron. The functional (\ref{Jpol1}) is a result of the
optimization of the total adiabatic energy functional with respect
to the polarization of the medium thereby exhibiting an effective
self-interaction of the electron.\cite{appel,rashba,YNGpol}
Correspondingly, the $\Upol$ term in Eq.~(\ref{Jpol1}) is known to
be ``made of'' two parts: $-\Upol=-\Upol_{el}+\Upol_{d}$, where
$\Upol_{el}$ represents the magnitude of the potential energy of
the electron in the polarization field and $\Upol_{d}$ the energy
required to create this polarization (``deformation energy'');
with the optimal polarization, $\Upol_{d}=\Upol_{el}/2$.

Both energy functionals (\ref{Jexc1}) and (\ref{Jpol1}) assume
that the electron and hole energies are measured from the band
edges.

It is convenient to factor out dependences  on physically relevant
combinations of parameters by introducing appropriate units of
energy and length. We will choose such units based on combinations
for the exciton Bohr radius and binding energy (effective Rydberg)
in 3$d$:
\begin{equation}\label{units}
\aB=2 \epsilon \hbar^2/m e^2, \ \ \ \Ry=e^2/2\epsilon \aB.
\end{equation}
For the exciton problem, Eq.~(\ref{Jexc1}), one uses
$\epsilon=\epinf$ in Eq.~(\ref{units}) and for the polaron
problem, Eq.~(\ref{Jpol1}), it is $\epsilon=\epstar$. We use
superscript indices ``exc'' and ``pol'' to distinguish between the
corresponding units (\ref{units}). With all the coordinates ($x$,
$y$ for the cylinder) measured in appropriate $\aB$, one arrives
at dimensionless energy functionals: $\Jexco=\Jexc/\Rys{exc}$ and
$\Jpolo=\Jpol/\Rys{pol}$, where
\begin{equation}
\Jfexco = \intr \dpsi - 2\intr \frac{\ppsir}{\disto{\vrr}}
\label{Jexc2}
\end{equation}
and
\begin{equation}
\Jfpolo = \frac{1}{2}\intr \dpsi - \introt \frac{\ppsiro \,
\ppsirt}{\disto{\vrro-\vrrt}}. \ \label{Jpol2}
\end{equation}
The dimensionless $\disto{\vrr}$ in Eqs.~(\ref{Jexc2}) and
(\ref{Jpol2}) features the same behavior as Eq.~(\ref{dist}) but
with $R$ replaced by the corresponding $\Ro=R/\aB$.

The global minima of Eqs.~(\ref{Jexc2}) and (\ref{Jpol2}):
$-\bexco$ and $-\bpolo$, respectively, would determine the
dimensionless binding energies. As units (\ref{units}) already
establish the scaling rules, in what follows we will be comparing
$\bexco$ and $\bpolo$ at the same values of $\Ro$. The ratio
$\bpolo/\bexco$ would have a direct physical meaning of maximum
achievable when $\eps \gg \epinf$ and $\epstar \simeq \epinf$ in
Eq.~(\ref{epstar}). As we mentioned earlier, this can be a typical
situation for many polar solvents.

Before proceeding with the analysis for a cylindrical surface, we
recall in more detail benchmarks known for isotropic
$d$-dimensional systems ($\disto{\vrr}=|\vrr|=r$) briefly
described in the Introduction. The exact isotropic excitonic
ground state $\psi (r)$ corresponding to Eq.~(\ref{Jexc2}) is
given by the solution of the Schr\"{o}dinger equation
$$
-\bexco \psi  = -\frac{\partial^2 \psi}{\partial r^2} -
\frac{(d-1)}{r}\,\frac{\partial \psi}{\partial r} - \frac{2}{r}\,
\psi,
$$
yielding well-known
\begin{equation}\label{ddim}
\psi (r) \propto \exp\left(-\frac{2r}{d-1}\right), \ \ \
\bexco=\frac{4}{(d-1)^2}.
\end{equation}
With our choice of units, $\bexco=1$ in 3$d$.

The polaronic ground-sate wave functions and energies
corresponding to Eq.~(\ref{Jpol2}) are known from variational
calculations for $d$-dimensional systems. It is customary in the
polaronic literature to express energies in terms of the coupling
constant $\alpha_{c}=\left(m e^4 /2\epsilon^{*2}\hbar^3 \omega
\right)^{1/2}$ and phonon frequency $\omega$. Note that the
combination $\alpha_{c}^2 \,\hbar\omega$ appearing in results for
strong-coupling (adiabatic) polarons corresponds to $2\Rys{pol}$
as defined in Eq.~(\ref{units}). Thus well-known Pekar's
result\cite{pekar1} for the 3$d$ polaron translates into
$\bpolo\simeq 0.218$ and into the same magnitude of the ratio
$\bpolo/\bexco$ in 3$d$.

\begin{figure}
\includegraphics[scale=0.7]{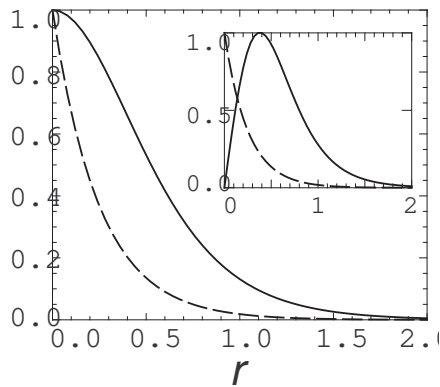}
\caption{\label{WF2d}The spatial distribution of unnormalized
density $\left|\psi (r)\right|^{2}$ for the exciton (dashed line)
and the polaron (solid line) in 2$d$. The polaron wave function is
as per Eq.~(\ref{my2d}). The inset shows the corresponding spatial
distribution $U(r)$ of the potential energy in $U=\int U(r) dr$.}
\end{figure}

The 2$d$ polaron has been studied in great detail in
Ref.~\onlinecite{pol2d} with the best result of $\bpolo\simeq
0.809$ for the adiabatic case achieved with Pekar-type trial wave
functions. Since $\bexco=4$ in 2$d$, the ratio $\bpolo/\bexco
\simeq 0.202$, only slightly smaller than in 3$d$. The 2$d$ case
represents the limit $R \rightarrow \infty$ for a cylindrical
surface and is particularly important for our analysis. We have
looked at simpler one-parametric trial wave functions that would
make a good representation of the 2$d$ adiabatic polaron and found
that the wave function
\begin{equation}\label{my2d}
\psi (r) \propto \frac{1}{\cosh\,(\al r)}
\end{equation}
with $\al \simeq 1.674$ yields a very good optimization for the
energy: $\bpolo \simeq 0.804$, quite accurate for our purposes.
Figure \ref{WF2d} compares the spatial structure of the 2$d$
exciton and polaron.

As both exciton and polaron binding energies diverge in pure 1$d$,
the following comparison might be of a dubious nature but is still
interesting. Specifically, Ref.~\onlinecite{pol1d} discussed the
calculation of the 1$d$ polaron in terms of the renormalized
coupling constant $\alpha_{c}^{\,\prime}=\alpha_{c}/(d-1)$, where
$d \rightarrow 1$. The best variational result achieved for the
adiabatic polaron was $\bpol \simeq 0.333 \, (\alpha_{c}^{\,\prime
2}\hbar\omega)$. As the diverging $d$-dependence in this
expression is the same $(d-1)^{-2}$ as in the exciton case
(\ref{ddim}), the ratio of the binding energies in 1$d$ could then
be interpreted as $\bpolo/\bexco \simeq 0.167$.

\section{Variational analysis for a cylinder}

When on a cylindrical surface, both exciton and polaron ground
states need to be determined numerically. Our variational analysis
of the energy functionals (\ref{Jexc2}) and (\ref{Jpol2}) has been
performed on the following classes of the trial wave
functions.\footnote{All numerical integrations and optimizations
have been done by using IMSL numerical libraries as provided with
the PV-WAVE Advantage package, http://www.vni.com.}  For the
exciton problem we have used three-parametric ($\al$, $\be$ and
$\ga$) wave functions
\begin{equation}\label{varexc}
\psi (x,y) \propto \exp\left[-\left(\al^2 x^2 + \be^2 \yo^2 +
\ga^2 \right)^{1/2}\right].
\end{equation}
The polaron problem is much more computation time demanding and we
chose two-parametric ($\al$ and $\be$) wave functions
\begin{equation}\label{varpol}
\psi (x,y) \propto \frac{1}{\cosh\left(\al^2 x^2 + \be^2
\yo^2\right)^{1/2}}.
\end{equation}
The functional dependences in Eqs.~(\ref{varexc}) and
(\ref{varpol}) are such that they can recover, in the limit of
$\Ro \rightarrow \infty$, wave functions (\ref{ddim}) and
(\ref{my2d}) found for the 2$d$ systems -- similarly to the
earlier exciton calculations.\cite{kostov,pedersen}

We explored two choices for the effective coordinate $\yo$ in
Eqs.~(\ref{varexc}) and (\ref{varpol}): ``arc-based'' (as in
Ref.~\onlinecite{pedersen})
\begin{equation}\label{arc}
\yo=y,
\end{equation}
$-\pi \Ro < y < \pi \Ro$, and ``chord-based'' (as in
Ref.~\onlinecite{kostov})
\begin{equation}\label{chord}
\yo=2\Ro\sin\left(y/2\Ro\right).
\end{equation}
Both choices can be thought of as respectively $n \gg 1$ and $n=1$
limits of more general
$$
\yo=\pi \Ro
\left\{\frac{2}{\pi}\sin\left[\frac{\pi}{2}\left(\frac{y}{\pi \Ro}
\right)^{n}\, \right] \right\}^{1/n}
$$
that could be used in future refinements as being more flexible in
terms of the shape of the wave function periodic in the
circumferential direction. In this paper we resorted to just
choosing the best results among obtained with Eqs.~(\ref{arc}) and
(\ref{chord}).

Wave functions (\ref{varexc}) and (\ref{varpol}) feature two
parameters $\al$ and $\be$ having the meaning of inverse lengths
thereby explicitly allowing for anisotropy of the wave function
extent in the axial and circumferential directions.\cite{pedersen}
What we will later be referring to as quasi-1$d$ results
corresponds to $\be=0$ when the wave functions are uniform around
the cylinder circumference.

\begin{figure}
\includegraphics[scale=0.7]{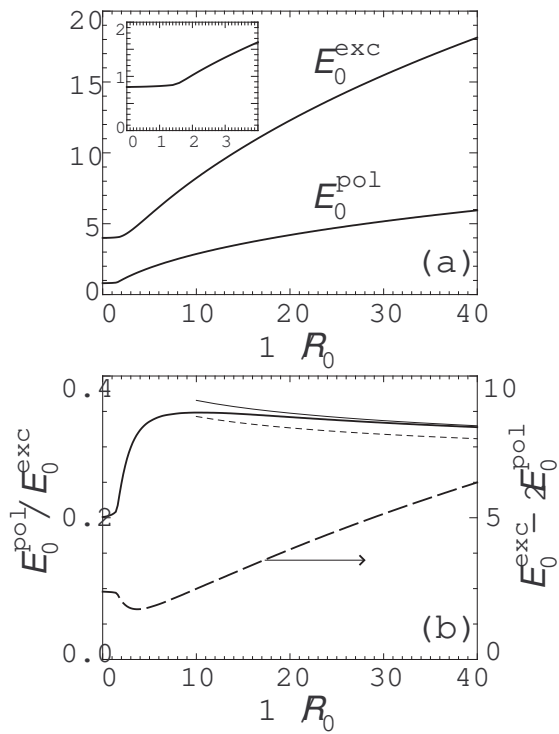}
\caption{\label{Energies}(a) The dimensionless exciton and polaron
binding energies as functions of the dimensionless inverse
cylinder radius $1/\Ro$. The inset shows the behavior of the
polaron binding energy in more detail at small $1/\Ro$. (b) The
ratio of the binding energies $\bpolo/\bexco$ of the excitations
on a cylinder is shown as a thick solid line. A thin solid line
displays this ratio as it is obtained in the quasi-1$d$
calculation with the effective tube potential (\ref{Vtube}). A
thin short-dash line shows the quasi-1$d$ result for this ratio if
the effective interaction, instead, was that of a quantum wire,
Eq.~(\ref{Vwire}). A thick long-dash line displays the
dimensionless activation energy $\bexco-2\bpolo$.}
\end{figure}

Main quantitative results of this paper are displayed in Figure
\ref{Energies} showing the optimized variational outputs for the
exciton and polaron binding energies as well as their ratio
$\bpolo/\bexco$. Our exciton data is very close to results of
Refs.~\onlinecite{kostov} and \onlinecite{pedersen} where the
reader can find extensive discussions.

Just as in the exciton case, the polaron binding energy exhibits
very little change from its 2$d$ value ($\simeq 0.8$) due to the
curvature up to $\Ro \sim 1$ (see the inset in
Fig.~\ref{Energies}(a)) where it starts rising, relatively earlier
and more rapidly than the exciton dependence. This immediately
translates into a substantial increase of the ratio
$\bpolo/\bexco$ with decreasing $\Ro$ as shown in
Fig.~\ref{Energies}(b). The maximum of the ratio $\simeq 0.35$ is
achieved in the region of $1/\Ro \sim 10$, after which the ratio
starts slowly decreasing with $1/\Ro$. While, of course, larger
ratios $\bpolo/\bexco$ lead to larger relative reductions of the
effective activation energy $\bexco-2\bpolo$, it is quite
interesting that the \textit{absolute} value of this activation
energy, also shown in Fig.~\ref{Energies}(b), exhibits a
non-monotonic dependence on $1/\Ro$. A region around the minimum
of this curve indicates specific tube sizes where the activation
energy would be at its lowest.

Our variational results have shown that the the dependence of
$\bpolo$ on $1/\Ro$ collapses onto the corresponding quasi-1$d$
curve practically right away after the onset of a substantial rise
in the binding. This is different from the exciton case where
deviations from the quasi-1$d$ behavior persist all the way into
the region of $1/\Ro > 20$. In other words, the polaron spreads
uniformly around the cylinder at much smaller curvatures than the
exciton does. We note that the quasi-1$d$ variational results for
the polaron binding using trial wave functions (\ref{varpol}) and
(\ref{varexc}) differ very little and are in good agreement with
our analysis in Ref.~\onlinecite{YNGpol} where no assumptions have
been made about the wave function shape and the nonlinear
optimizing equation has been solved numerically. The thin solid
line in Fig.~\ref{Energies}(b) shows the ratio $\bpolo/\bexco$
using quasi-1$d$ results for the binding energies and its
deviation from the variational result for a cylinder is entirely
due to underestimation of the exciton binding.

As with all variational calculations, we, of course, cannot
exclude that some details in the results shown in
Fig.~\ref{Energies} may undergo slight modifications upon further
improvements of variational wave functions. Importantly, possible
improvements would be inconsequential for our main observations of
a non-monotonic dependence of the ratio $\bpolo/\bexco$ on $1/\Ro$
and of the magnitude of the ratio reaching values well above the
2$d$ value of $\simeq 0.2$. We found that a rise of the ratio
above $0.3$ is obtainable even if the polaron wave functions are
not specifically optimized for a range of given $\Ro$ but the 2$d$
optimal values of $\al=\be\simeq 1.67$ are used. For very small
tubes with $1/\Ro \gtrsim 20$, results of exact numerical
calculations in Ref.~\onlinecite{YNGpol} complement the picture.

\begin{figure}
\includegraphics[scale=0.75]{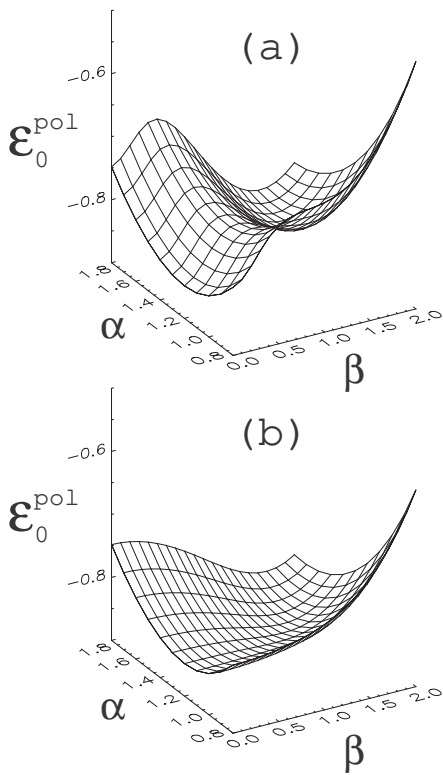}
\caption{\label{Landscapes}The behavior of the functional
(\ref{Jpol2}) at $\Ro=0.65$ with the trial wave functions
(\ref{varpol}) as a function of the variational parameters $\al$
and $\be$ for two choices of the effective coordinate $\yo$: (a)
as in Eq.~(\ref{arc}) and (b) as in Eq.~(\ref{chord}).}
\end{figure}

The transitional region of $\Ro \sim 1$ is likely to be especially
sensitive to the choice of trial wave functions. So
Ref.~\onlinecite{kostov} reported a few-percent improvement for
the exciton binding energy in the region of $1 \lesssim 1/\Ro
\lesssim 2.5$ with certain trial functions. Similar improvements
could perhaps be found for the polaron binding energies. Figure
\ref{Landscapes} illustrates the behavior of the polaron
functional (\ref{Jpol2}) in the transitional region, at
$\Ro=0.65$, as a function of the variational parameters $\al$ and
$\be$. A curious feature of the ``landscape'' in
Fig.~\ref{Landscapes}(a) is a clear coexistence of two minima, one
corresponding to a polaronic state that is uniformly distributed
(delocalized) around the cylinder circumference, and the other
where the circumferential distribution is non-uniform. While this
appears as an interesting possibility, it could as well be an
artefact of a restricted nature of a specific class of the trial
wave functions, compare, e.g., visually to the landscape of
Fig.~\ref{Landscapes}(b). A much more careful study would be
needed to explore the possibility of coexisting polaronic states
requiring an analysis of actual adiabatic potential surfaces. An
example of an analogous analysis can be found in
Ref.~\onlinecite{YNG86}, where we proved a coexistence of
different polarons in certain quasi-1$d$ systems.

\section{Discussion}

Both in this paper and in Ref.~\onlinecite{YNGpol} we have shown
that the strong polaronic effect occurring in sluggish polar
environments substantially affects the relative energetics of
closely-bound and well-separated electron-hole pairs on a
cylindrical surface of nanotubes. The binding energy $\bpol$ of an
individual polaron can, in principle, reach as much as about 0.35
of the binding energy $\bexc$ of the exciton. This would translate
into a reduction of the activation energy $\bexc - 2\bpol$ for
exciton dissociation by a factor of about 3 from the value $\bexc$
it would have in a non-polar environment with the same value of
the high-frequency dielectric constant $\epinf$. Note that we have
not found any additional energy barriers between the exciton and
distant polaron-pair states.\cite{YNGpol} One should expect that
enhanced separation of charges and a corresponding luminescence
quenching would then result, e.g., in experiments probing electric
field effects on the luminescence. Needless to say that additional
factors can make the reduction magnitude we discussed smaller,
see, e.g., a comparison in Ref.~\onlinecite{YNGpol} of cases with
different ``electrostatic conditions''.

Polar liquid environments such as many common solvents may be good
candidates to provide conditions necessary for the strong
polaronic effect as they can exhibit both high values of the
static dielectric constant $\eps$ and a relatively slow response
of the orientational polarization (longitudinal relaxation times
can be on the order of 1 ps and longer)\cite{fawcett,YNGpol}.
Quite befittingly, there is an intense ongoing research effort on
various applications of nanotubes in contact with such
environments and we hope that direct experimental verifications of
our conclusions would be possible. We note that a comprehensive
mapping of luminescence versus absorption spectra of individual
SWNTs was in fact achieved in aqueous
suspensions.\cite{micelle1,micelle2} Interestingly, numerical
estimates in Ref.~\onlinecite{pedersen} indicate that the
dimensionless radius $\Ro$ for a range of SWNTs may be close to
0.1, which corresponds to the region of maximum $\bpolo/\bexco$
ratios in Fig.~\ref{Energies}(b).

We have demonstrated that the ratio of the binding energies
$\bpol/\bexc$ has a non-monotonic dependence on the cylinder
curvature $1/R$. As argued in the Introduction, our particular
observation for the cylindrical geometry can be a manifestation of
a more general crossover effect that would be common for other
structures in 3$d$ polar media when the increasing confinement of
the electron motion causes a ``transition'' between
$d$-dimensional systems, such as between 3$d$ and 2$d$ (quantum
wells) or between 3$d$ and 1$d$ (quantum wires). Basically, the
origin of this effect can be related to the fact that the spatial
extent of an unconfined polaron is larger than the size of an
unconfined exciton (see Fig.~\ref{WF2d} for the 2$d$ case) thereby
making the polaron ``respond'' to the initially increasing
confinement in a more pronounced way than the exciton. Only after
the exciton experiences a fuller effect of the confinement, the
ratio of the binding energies starts decreasing. Elaborating more
on this idea, the inset in Fig.~\ref{WF2d} shows the spatial
distribution of the potential energy terms $U$: $\mathcal{E}=K-U$,
for unconfined excitations. It is evident that the
``longer-range'' contributions to the polaron potential energy are
relatively more important than for the exciton. That is why the
effect of the increasing confinement on the polaron is initially
relatively stronger. In the particular case of the cylindrical
geometry, the curvature changes remote physical distances (chords
instead of arcs, Eq.~(\ref{dist})) more than it does close
distances.

A meaningful parallel can be drawn with the behavior of the ratio
$U/K$ of the potential and kinetic energy terms in confined
systems. Virial theorem for the Coulombically bound states in
unconfined $d$-dimensional systems states that $U/K=2$
independently of $d$, which, of course, also follows directly from
scaling of both functionals (\ref{Jexc1}) and (\ref{Jpol1})
provided that $\dist{\vrr}=|\vrr|=r$. Particularly, polarons in
such systems are known\cite{pekarbook,rashba,appel} to satisfy the
following ratios for various energy terms: \
$\bpol\!:\!\Kpol:\!\Upol:\!\Upol_{el}\!:\Upol_{d} =
1\!:\!1:\!2:\!4:\!2$. Some of these relationships are violated in
confined systems. As studied in Ref.~\onlinecite{zhang}, the
virial theorem ratio $\Uexc/\Kexc$ for excitons in quantum wells
and quantum wires is larger than 2 and, in fact, a non-monotonic
dependence of $\Uexc/\Kexc$ has been demonstrated for quantum
wells transitioning between 3$d$ and 2$d$ limits. We have found a
non-monotonic behavior of the ratio $U/K$ for both polarons and
excitons on a cylinder as a function of the curvature $1/R$. In
agreement with our qualitative arguments, at smaller $1/R$, this
ratio grows much faster for the polaron than for the exciton. At
large curvatures, however, the trend is reversed and the exciton
has larger ratios $U/K$ than the polaron.

\begin{figure}
\includegraphics[scale=0.7]{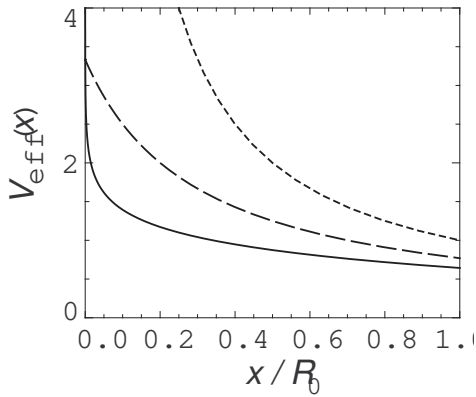}
\caption{\label{Potentials}Distance dependence of the effective
1$d$ Coulomb potentials for charges on a cylinder,
Eq.~(\ref{Vtube}), solid line, and in a quantum wire,
Eq.~(\ref{Vwire}), long-dash line, in comparison with the original
Coulomb interaction (\ref{V0}), short-dash line.}
\end{figure}

It is useful to continue a qualitative reasoning by discussing the
quasi-1$d$ limit of our results, that is, the case of stronger but
still finite degrees of confinement. One can then use the notion
of the effective Coulomb potentials,\cite{haugbook,zhang} here as
a function of the 1$d$ (axial) distance $x$. (The effective
Coulomb potentials for 2$d$ can be similarly
introduced.\cite{zhang}) Figure \ref{Potentials} compares three
potentials in units such that at very large distances the
potentials behave as
\begin{equation}\label{V0}
\Veff=\Ro/x.
\end{equation}
In this limit the electron wave function is delocalized around the
tube circumference and the effective interaction becomes that of
rings of charge given by
\begin{equation}\label{Vtube}
\Veff=\frac{2}{\pi\left[(x/\Ro)^2+4\right]^{1/2}}\,
K\left[\frac{4}{(x/\Ro)^2+4} \right],
\end{equation}
where $ K(m)=\int_{0}^{\pi/2} \left(1-m\sin^2 \theta\right)^{-1/2}
d\theta $ is a complete elliptic integral of the first kind. If,
instead of a tube, we dealt with a quantum wire, the electron wave
function would be delocalized throughout the cross-section of the
wire with the effective interaction approximated as\cite{haugbook}
\begin{equation}\label{Vwire}
\Veff=\frac{1}{(x/\Ro)+0.3}.
\end{equation}

Both effective tube (\ref{Vtube}) and wire (\ref{Vwire}) 1$d$
potentials feature modifications of the shorter-range interaction
from the original Coulomb (\ref{V0}) due to the transverse spread
of wave functions thereby eliminating a pure 1$d$ divergence of
the ground states for excitons and polarons. As we discussed
above, the role of the longer-distance interactions is more
important for the polaron than for the exciton. Since the relative
modification of the original Coulomb to the effective potentials
is increasing towards shorter distances, it is then clear that the
ratio $\bpol/\bexc$ in systems with modified interactions should
be larger than values around $0.2$ in systems with the pure
Coulomb interaction. Moreover, following the same logic, one
should expect that the larger modification is from the original
Coulomb distance dependence the larger would be $\bpol/\bexc$
ratios. Figure \ref{Potentials} shows that deviations from the
Coulomb dependence for the effective wire potential are smaller
than for the tube potential (over a relevant spatial range). We
have performed a quasi-1$d$ variational optimization of the
polaron and exciton binding energies with the tube potential
(\ref{Vtube}). The resulting ratios $\bpolo/\bexco$ are shown in
Fig.~\ref{Energies}(b) with a short-dash thin line and indeed
smaller than the ratios calculated with the tube potential
(\ref{Vtube}), solid thin line in that figure.

Our qualitative arguments, while confirmed by specific
calculations, do not appear to be restricted to this specific
situation. We therefore believe that findings of larger magnitudes
of the ratio $\bpol/\bexc$ and of a non-monotonic dependence of
this ratio on the degree of confinement represent a crossover
effect that can be common to semiconductor nanostructures in 3$d$
polar environments.  Further calculations with different
structures are needed to validate this conjecture and evaluate its
quantitative aspects, including in non-uniform polar environments.

Another important venue for future research is an assessment of
the activation energy for exciton dissociation in confined
semiconductors with unequal electron and hole masses, $m_{e}\neq
m_{h}$. In this case electron- and hole-polarons have different
binding energies and, in addition, the exciton itself can cause an
adiabatic polarization of the environment.\cite{pekar2,pekar3} The
corresponding ``polaronic'' corrections to the exciton binding in
quantum-well wires have, e.g., been studied in
Ref.~\onlinecite{degani87}. From a general standpoint, one would
also like to extend the analysis of the effects of polar media on
the dissociation of confined excitons to the
intermediate-coupling\cite{froreview,polarons1,appel} case.

\acknowledgments

We are deeply grateful to V.~M.~Agranovich for many useful
discussions. This study contributes to a project supported by the
Collaborative U.~T.~Dallas -- SPRING Research and Nanotechnology
Transfer Program. The work of SOC was supported by the University
of Texas at Dallas Summer Research Program.
\newpage

\bibliography{polar}

\end{document}